\documentclass[12pt,usenames,dvipsnames]{article}

\usepackage{latexsym}
\usepackage{amssymb,amsfonts,amsmath}
\usepackage{graphicx} 
\usepackage{indentfirst}
\usepackage{bbm}
\usepackage{amssymb}
\usepackage{verbatim}
\usepackage{amsmath, amsthm,amssymb}
\usepackage{mathrsfs}
\usepackage{hyperref}
\usepackage{amsfonts}
\usepackage{dsfont}
\usepackage{cite}
\usepackage{xcolor}
\usepackage{enumerate}
\usepackage{cleveref}

\parskip=\medskipamount

\newcommand {\cD}{{\cal D}}
\newcommand {\cE}{{\cal E}}

\newcommand {\cK}{{\cal K}}
\newcommand {\cL}{{\cal L}}
\newcommand {\cM}{{\cal M}}
\newcommand {\cN}{{\cal N}}

\newcommand {\cQ}{{\cal Q}}
\newcommand {\cR}{{\cal R}}

\newcommand {\cT}{{\cal T}}
\newcommand {\cU}{{\cal U}}

\def\a{\alpha}

\def\b{\beta}

\def\d{\delta}

\def\f{\phi}
\def\g{\gamma}
\def\G{\Gamma}

\def\m{\mu}

\def\o{\omega}

\def\q{\theta}

\def\s{\sigma}

\def\D{\Delta}

\def\J{\Psi}
\def\L{\Lambda}
\def\O{\Omega}

\def\rd{{\rm d}}
\def\ri{{\rm i}}

\newcommand{\ad}{{\dot{\alpha}}}                           %
\newcommand{\bd}{{\dot{\beta}}}                            %
\newcommand{\ve}{\varepsilon}                            %
\newcommand{\cDB}{{\bar\cD}}                            %

\renewcommand{\aa}{{\a\ad}}
\newcommand{\bb}{{\b\bd}}
\newcommand{\pa}{\partial}                           %
\newcommand{\hf}{\frac12}

\newcommand{\be}{\begin{equation}}
\newcommand{\ee}{\end{equation}}
\newcommand{\bea}{\begin{eqnarray}}
\newcommand{\eea}{\end{eqnarray}}
\newcommand{\non}{\nonumber}

\newcommand{\dsC}{{\mathbb C}}

\def\double #1{#1{\hbox{\kern-2pt $#1$}}}

\newcommand{\gd}{{\dot\g}}
\newcommand{\dd}{{\dot\d}}

\newcommand{\sba}{{\bar{\s}}}

\newcommand{\OI}{{\overline{I}}}
\newcommand{\OJ}{{\overline{J}}}

\newcommand{\UI}{{\underline{I}}}
\newcommand{\UJ}{{\underline{J}}}

\newif\ifdtup

\newcommand{\bsubeq}{\begin{subequations}}
\newcommand{\esubeq}{\end{subequations}}

\numberwithin{equation}{section}

\newcommand{\sSU}{\mathsf{SU}}
\newcommand{\sSL}{\mathsf{SL}}

\newcommand{\sU}{\mathsf{U}}

\begin{document}

\begin{titlepage}
\begin{flushright}
November, 2026 
\end{flushright}
\vspace{5mm}

\begin{center}
{\Large \bf The logarithm construction and (super) Weyl anomalies}
\end{center}

\begin{center}

{\bf Sergei M. Kuzenko} \\
\vspace{5mm}

\footnotesize{
{\it Department of Physics M013, The University of Western Australia\\
35 Stirling Highway, Perth W.A. 6009, Australia}}  
~\\

\vspace{2mm}

\end{center}

\begin{abstract}
\baselineskip=14pt
We apply the logarithm construction of higher-derivative conformal primary (super) fields proposed by Butter et al. to derive simple Wess-Zumino actions for: (i)  Weyl anomalies in arbitrary even dimensions; (ii) $\mathcal{N} =1$ and $\mathcal{N}=2$ super-Weyl anomalies in four dimensions; and (iii) $(p,q)$ super-Weyl anomalies in two dimensions, with $p,q \leq 2$.
\end{abstract}
\vspace{5mm}

\vfill

\vfill
\end{titlepage}

\newpage
\renewcommand{\thefootnote}{\arabic{footnote}}
\setcounter{footnote}{0}

\tableofcontents{}
\vspace{1cm}
\bigskip\hrule

\allowdisplaybreaks


\section{Introduction} \label{Section1}

For Weyl (or conformal) anomalies \cite{Duff:1977ay, Duff:1993wm} in arbitrary ($d= 2n$) spacetime dimensions, a complete geometric description was given by Deser and Schwimmer \cite{DS}, several years after the cohomological analysis of these anomalies was provided by Bonora, Pasti and Bregola \cite{Bonora:1985cq}.
In classically Weyl-invariant theories, the anomalous trace of the energy-momentum tensor has the general form\footnote{We ignore an overall numerical factor $(4\pi)^{d/2}$  and a sign factor $(-1)^{d/2}  $ in the $a$-term, which are often included in the definition of the Weyl anomaly. In what follows, we will also omit scheme-dependent total derivative terms. }
\bea
\langle T\rangle = \sum_i c_i I_i +a E_d + \text{total derivative}~.
\label{trace}
\eea
Here the first term represents the so-called type B anomaly. It is given by a linear combination of conformal primary fields $I_i$, of Weyl weight $d$,  which are constructed from the Weyl tensor and its covariant derivatives. The second term in \eqref{trace}  describes the type A anomaly and involves the Euler scalar $E_d$. A comprehensive review of the Weyl anomalies in two and four dimensions can be found in \cite{Duff:1993wm}.
The structure of Weyl anomalies in six dimensions was studied in the late 1990s \cite{Arakelian:1995ye,  KMM, Bastianelli:2000hi};
see \cite{Casarin:2024qdn} for recent results and a complete list of references. For a recent cohomological analysis of Weyl anomalies in diverse dimensions see
\cite{Boulanger:2007ab}.

In this paper, we present simple Wess-Zumino actions for the Weyl anomalies in arbitrary dimensions and for their supersymmetric extensions: 
$\cN=1$ and $\cN=2$ super-Weyl anomalies in four dimensions \cite{BPT,BK86,K13} and their two-dimensional cousins. Our approach is based 
on applying the logarithm construction of primary higher-derivative (super)fields proposed in \cite{Butter:2013lta}.
For completeness, it is instructive to provide a brief review of the logarithm construction
in the non-supersymmetric case, where one  builds on the formulation for conformal gravity 
in $d \geq 3$ spacetime dimensions following \cite{BKNT-M1,BKNT} (see also a recent review \cite{Kuzenko:2026vir}).
Of course, conformal gravity has been discussed elsewhere in many places, including the comprehensive textbook \cite{FVP}. 
However, the role of conformally covariant derivatives $\nabla_a$ were not discussed/emphasised  
in the original papers on conformal (super)gravity \cite{KTvN,KTvN2}. 
Their properties were described several years later by Kugo and Uehara \cite{Kugo:1983mv}. These operators are of crucial importance for the logarithm construction. 

Consider a curved spacetime $(\cM^d, e_m{}^a , \o_m{}^{\underline a})$, with $e^a =\rd x^m e_m{}^a$ and 
$\o^{\underline{a}} = (\hat{\o}^{ab},\mathfrak{b},\mathfrak{f}_a)=\rd x^{m} \o_m{}^{\underline{a}}$ the connection one-forms 
of conformal gravity, see Appendix \ref{AppendixA} for the technical details. 
Conformal gravity is the gauge theory of the conformal algebra $\mathfrak{so}(d,2)$.  It is described by the conformally covariant derivatives 
\bea
\nabla_a 
= e_a{}^m \partial_m - \hf \hat{\omega}_a{}^{bc} M_{bc} - {\mathfrak b}_a \mathbb D - \mathfrak{f}_{ab} K^b ~,
\eea
involving the Lorentz ($M_{bc}$), dilatation ($\mathbb{D}$) and special conformal  ($K^b$) generators. 
For $d>3$ they obey the commutation relations 
\begin{align}
[\nabla_a,\nabla_b]=-\frac{1}{2}C_{abcd}M^{cd}-\frac{1}{2(d-3)}\nabla^dC_{abcd}K^c~,
\label{Cotton}
\end{align}
with $C_{abcd}$ the Weyl tensor.

Let $\Phi$ be a conformal primary field  on $\cM^d$  of Weyl weight (or dimension) zero, 
\bea 
K^b \Phi =0~, \qquad {\mathbb D} \Phi =0~.
\eea
On any spacetime $\cM^d$ there exists a conformally covariant operator (belonging to the family of  GJMS operator \cite{GJMS, Graham, Gover})  of the form\footnote{This operator is not unique beyond four dimensions. In six dimensions, there is a two-parameter family of operators \cite{Arakelian:1995ye}. It is assumed that $\Delta_d$ has no potential-like term, that is 
$\Delta_d 1 =0$.} 
\bea
\Delta_d = \square_{\rm c}^{d/2} +\text{lower order terms}~, \qquad \square_{\rm c} = \nabla^a \nabla_a~,
\label{Delta}
\eea
which takes $\Phi$ to a primary field of Weyl weight $d$, 
\bea
K^b \Delta_d \Phi=0~, \qquad {\mathbb D}\Delta_d \Phi = d\Delta_d \Phi~.
\eea
The ellipsis in \eqref{Delta} stands for lower-order terms involving the Weyl tensor and its covariant derivatives. For $d\leq6$ these operators are: 
\begin{subequations} 
\begin{align} 
\Delta_2 =& \square_{\rm c}~, \\
\Delta_4 =& \square_{\rm c}^2 ~, \label{Paneitz}\\
\Delta_6 =& \square_{\rm c}^3 - \frac 83 (\nabla^b \nabla^d C_{abcd}) \nabla^a \nabla^c~.\label{Box3}
\end{align}
\end{subequations} 
The operator $\Delta_4$ was discovered in \cite{FT1982, Paneitz, Riegert}, while $\Delta_6$ 
was constructed in \cite{Branson, Wunsch} (see also \cite{Arakelian:1995ye,  KMM}).\footnote{Strictly speaking, the degauged versions of the operators \eqref{Paneitz} and \eqref{Box3} were derived in 
 \cite{FT1982, Paneitz, Riegert} and \cite{Branson, Wunsch}, respectively. In terms of the conformally covariant derivatives, these operators 
 appeared in \cite{Butter:2013lta} and  \cite{BKNT}, respectively.}
Given two  primary dimensionless scalar fields $\Phi_1$ and $\Phi_2$, it holds that 
\bea
\int \rd^d x \, e\, \Phi_1 \Delta_d \Phi_2 = \int \rd^d x \, e\, \Phi_2 \Delta_d \Phi_1~, \qquad e = {\rm det} (e_m{}^a)~.
\eea
This functional is conformally invariant, for any $\Phi_1$ and $\Phi_2$.
Since $\Delta_d 1 =0$, we observe that  $e\, \Delta_d \Phi$ is a total derivative, and therefore
\bea
\int \rd^d x \, e\,  \Delta_d \Phi = 0~,
\eea
assuming the filed vanishes at infinity.

Let $\Psi$ be a conformal primary real scalar field of Weyl weight $w \neq 0$, 
\bea
K^b \Psi=0~, \qquad {\mathbb D}\Psi = w \Psi ~.
\label{Psi}
\eea
We assume $\Psi $ to be nowhere vanishing, $\Psi (x) >0$ at every spacetime point $x$. Then $\Delta_d \ln \Psi $ is conformal primary, 
\bea
K^b\Delta_d \ln \Psi =0~, \qquad {\mathbb D}\ln \Psi = d \Delta_d \ln \Psi~.
\label{PrimaryLog}
\eea
This result was established in \cite{Butter:2013lta} in the $d=4$ case, and in \cite{BKNT} for $d=6$. 
Although we have not done explicit calculations beyond six dimensions, we conjecture eq. \eqref{PrimaryLog} to be true for $d=2n>6$.

In this paper we will show how to apply the primary field $\Delta_d \ln \Psi$ to construct Wess-Zumino actions for Weyl anomalies. Supersymmetric generalisations will be also described. 


\section{Wess-Zumino action for Weyl anomalies}  \label{Section2}

In order to relate the discussion in Section \ref{Section1} to the standard formalism of general relativity, we should perform the procedure of degauging from 
the conformally covariant derivatives $\nabla_a$ to torsion-free Lorentz covariant derivatives $\cD_a$. This procedure is described in Appendix \ref{AppendixA}. 
Upon degauging, we obtain
\bea
\Delta_d \Phi = \widehat{\Delta}_d \Phi~, \qquad  \widehat{\Delta}_d = (\cD^a \cD_a)^{d/2} + \text{lower order terms}~,
\eea
where the operator $\widehat{\Delta}_d $ has no potential-like term, that is $\widehat{\Delta}_d 1=0$.
We also obtain 
\bea
\Delta_d \ln \Psi = \widehat{\Delta}_d \ln \Psi + w \cQ_d~.
\eea
Here $\cQ_d$ is a composite scalar field with the following properties: (i) $\cQ_d$ is independent of $\Psi$; (ii) $\cQ_d$ is constructed from the background gravitational field; and (iii) modulo numerical factor, $\cQ_d$ is the $\cQ$-curvature
\cite{Branson}.

For example, the degauging of  $\Delta_4 \ln \Psi = \square_{\rm c} \square_{\rm c} \ln \Psi$ leads to 
\begin{align} \label{D_00}
 \Delta_4   \ln \Psi = \Big(\square \square 
 -  \stackrel{\longrightarrow}{\mathcal{D}^a} \big(
	2 {R}_{ab} 
		- \frac{2}{3}\eta_{ab} {R} \big) \mathcal{D}^b \Big)\ln \Psi
		+ w \Big( \frac 16 \Box R - \frac 12 R^{ab}R_{ab} + \frac 16 R^2\Big) ~,
\end{align}
with $\square =\cD^a \cD_a$.
It follows that the primary field $ C^{abcd} C_{abcd} + 4 w^{-1}  \square_{\rm c} \square_{\rm c}  \ln \Psi $, introduced in   \cite{Butter:2013lta},
differs from the Gauss-Bonnet scalar
\bea
e^{-1} \cL_\chi=C^{abcd} C_{abcd} - 2 R^{ab} R_{ab} +\frac 23 R^2
\eea
by a total derivative. It should be pointed out that the first term in the right-hand side of \eqref{D_00} vanishes for $ \Psi = \text{const}$.
It should also be emphasised that 
\bea
\widehat{\Delta}_4 = \square \square 
  -  \stackrel{\longrightarrow}{\mathcal{D}^a} \big(
	2 {R}_{ab} 
		- \frac{2}{3}\eta_{ab} {R} \big) \mathcal{D}^b 
		\label{FTPR}	
\eea
is the Fradkin-Tseytllin-Paneitz-Riegert (FTPR) operator  \cite{FT1982, Paneitz, Riegert}.\footnote{In $d$ dimensions, the operator \eqref{Paneitz} is conformally invariant on the space of conformal primary fields $\Psi$ of Weyl weight $w =d/2 -2$. In other words,  $K^b \Delta_4 \Psi =0$
provided $\Psi$ the conformal properties \eqref{Psi} with $w=d/2 -2$.   Its degauged form was constructed by Paneitz \cite{Paneitz}.}

Upon degauging, the condition \eqref{Psi} turns into the following Weyl transformation law
\bea
\d_\sigma \Psi = w \sigma \Psi~.
\eea 
Here the parameter $\sigma $ generates the infinitesimal  Weyl transformation \eqref{WeylTr}.
Relation \eqref{PrimaryLog} implies the Weyl transformation law
\begin{subequations}
\bea
\d_\sigma \Delta_d \ln \Psi = d \sigma \Delta_d \ln \Psi ~,
\eea
and therefore 
\bea
 \delta_\sigma \widehat{\Delta}_d \ln \Psi = d \sigma \widehat{\Delta}_d \ln \Psi  +w \widehat{\Delta}_d \sigma~.
 \eea
\end{subequations}
We observe that the Weyl transformation law of $\cQ_d$ is 
\bea
\delta_\sigma \cQ_d = d \sigma \cQ_d - \widehat{\Delta}_d \sigma~.
\eea

Now let us come back to the Weyl anomaly \eqref{trace}. The above $d=4$ example indicates that the Euler scalar $E_d$ is related to the $\cQ$-curvature
\cite{Alexakis:2005ft, Alexakis} (see also \cite{Boulanger:2018rxo, Boulanger:2025oli})
as follows: 
\bea
E_d = \alpha_d \cQ_d + \sum_i \beta_i I_i +\text{total derivative}~,
\eea
for some numerical coefficients $\alpha_d$ and $\beta_i$.
Therefore, eq. \eqref{trace} can be recast as 
\bea
\langle T\rangle = 
\mathfrak{I}_d 
+ \tilde{a} \cQ_d + \text{total derivative}~, \qquad \mathfrak{I}_d = \sum_i \tilde{c}_i I_i ~,
\label{trace2}
\eea
for some numerical coefficients $\tilde a$ and $\tilde{c}_i $. By construction, $\mathfrak I$ is a primary field of dimension $d$,
\bea
\delta_\sigma \mathfrak{I}_d = d \sigma \mathfrak{I}_d ~.
\eea
The Weyl variation of the effective action is 
\bea
\delta_\sigma \G = - \int \rd^d x \, e\, \sigma \langle T\rangle~.
\eea
The overall minus sign here is due to our definition of the infinitesimal Weyl transformation \eqref{WeylTr} which 
implies that the spacetime metric $g_{mn}$ transforms as 
\bea
\d_\s g_{mn} = - 2\s g_{mn}~.
\eea

We proposed the following Wess-Zumino action: 
\begin{align}
 S_{\rm WZ} [e_m{}^a ; \ln \Psi] = &- \frac{1}{w} \int \rd^d x\, e\, \ln \Psi \Big\{  \mathfrak{I}_d + \frac{\tilde{a}}{w}  \Delta_d \ln \Psi\Big\} 
+ \frac{\tilde a}{2w^2}  \int \rd^d x\, e\, \ln \Psi \, \widehat{\Delta}_d \ln \Psi \non \\
 =&- \frac{1}{w} \int \rd^d x\, e\, \ln \Psi \Big\{  \mathfrak{I}_d + \tilde{a}  \cQ_d \Big\} 
- \frac{\tilde a}{2w^2}  \int \rd^d x\, e\, \ln \Psi \, \widehat{\Delta}_d \ln \Psi ~.
\label{WZaction}
\end{align}
A short calculation gives
\bea
\delta_\sigma S_{\rm WZ} [e_m{}^a ; \ln \Psi]  = - \int \rd^d x \, e\, \langle T\rangle~.
\eea
The Wess-Zumino consistency condition \cite{WZ-anomaly} is automatically satisfied. 
In the $d=4$ and $d=6 $ cases, the final expression  for $S_{\rm WZ}$ in \eqref{WZaction} is equivalent to the results obtained by the Armenian collaboration 
\cite{KMM}. 


\section{Wess-Zumino action for $\cN =1$ super Weyl anomalies} \label{Section3} 

In the superconformal case, our discussion will be more condensed than in Section \ref{Section2}. To describe $\cN=1$ conformal supergravity, we will make use of the $\cN=1$ conformal superspace formulation developed by Butter \cite{ButterN=1}, see \cite{KRTM1} for a detailed pedagogical review. A brief account of $\cN=1$ conformal superspace is given in Appendix \ref{AppendixB}. 

Let $\Phi$ be a primary chiral superfield of weight 0, 
\bea
K^B \Phi =0~, \qquad {\mathbb D}\Phi =0~, \qquad \bar \nabla_\ad \Phi =0~.
\eea
The $\cN=1$ superconformal analogue of the 
operator \eqref{Paneitz} was constructed in  \cite{Butter:2013lta}. 
It has the form
\begin{align}
\Delta \bar\Phi:= -\frac{1}{64} \bar\nabla^2 \nabla^2 \bar\nabla^2 \bar\Phi ~.
\end{align}
It turns out that $\Delta \bar\Phi$ is a primary chiral superfield of weight $+3$, 
\bea
K^B \Delta \bar\Phi =0~, \qquad {\mathbb D}\Delta \bar\Phi =3 \Delta \bar\Phi~, \qquad \bar \nabla_\ad \Delta \bar\Phi=0~.
\label{superPhi1}
\eea

Let $\Psi $ be a nowhere vanishing primary chiral superfield of weight $w\neq 0$, 
\bea
K^B \Psi =0~, \qquad {\mathbb D}\Psi =w \Psi~, \qquad \bar \nabla_\ad \Psi =0~.
\label{superPsi1}
\eea
It was shown in \cite{Butter:2013lta} that $\Delta \ln \bar\Psi$ is a primary chiral superfield,
\bea
K^B \Delta \ln \bar\Psi =0~, \qquad {\mathbb D}\Delta \ln \bar\Psi =3 \Delta \ln \bar\Psi~, \qquad \bar \nabla_\ad \Delta \ln \bar\Psi =0~.
\label{lnPsi1}
\eea

Upon degauging one obtains 
\begin{subequations}\label{eq:T1phib}
\begin{align}
\Delta \ln \bar\Psi &=
	\widehat{\Delta} \ln\bar\Psi + w \cQ~, 
	\end{align} 
	where we have introduced the operator
\begin{align} 
\widehat{\Delta} &:= 
	-\frac{1}{64} (\bar \cD^2 - 4 R) \Big(
	\cD^2 \bar \cD^2 
	+ 8 \cD^\alpha (G_{\alpha \ad} \bar \cD^\ad 
	)
	\Big)~,  
	\end{align}
	and the chiral superfield
	\begin{align}
\mathcal{Q} &:= - \frac{1}{4} (\bar\cD^2 - 4 R) \Big(
	 G^a G_a	+2 R \bar R  - \frac{1}{4} \cD^2 R
	\Big)~, \qquad \bar \cD_\ad \cQ =0~,
\end{align}
\end{subequations}
Here $\widehat \Delta $ is the $\cN=1$ supersymmetric generalisation of the  Fradkin-Tseytllin-Paneitz-Riegert operator \eqref{Paneitz}. The interpretation of $\cQ$ 
as the $\cN=1$ supersymmetric $\cQ$-curvature was given in \cite{Levy:2018xpu}.

Under a super-Weyl transformation \cite{HT, Siegel:1978fc} generated by a chiral parameter $\s$, $\bar \cD_\ad \s=0$, 
the covariant derivatives transform as 
\begin{subequations} 
\label{superweyl}
\bea
\d_\s \cD_\a &=& ( {\bar \s} - \hf \s)  \cD_\a + (\cD^\b \s) \, M_{\a \b}  ~, \\
\d_\s \bar \cD_\ad & = & (  \s -  \hf {\bar \s})
\bar \cD_\ad +  ( \bar \cD^\bd  {\bar \s} )  {\bar M}_{\ad \bd} ~,\\
\d_\s \cD_{\a\ad} &=& \hf( \s +\bar \s) \cD_{\a\ad} 
+\frac{\ri}{2} (\bar \cD_\ad \bar \s) \cD_\a + \frac{\ri}{2} ( \cD_\a  \s) \bar \cD_\ad \non \\
&& + (\cD^\b{}_\ad \s) M_{\a\b} + (\cD_\a{}^\bd \bar \s) \bar M_{\ad \bd}~.
\eea
\end{subequations}
The torsion tensors  transform
 as follows:
\begin{subequations} 
\bea
\d_\s R &=& 2\s R +\frac{1}{4} (\bar \cD^2 -4R ) \bar \s ~, \\
\d_\s G_{\a\ad} &=& \hf (\s +\bar \s) G_{\a\ad} +\ri \cD_{\a\ad} ( \s- \bar \s) ~, 
\label{s-WeylG}\\
\d_\s W_{\a\b\g} &=&\frac{3}{2} \s W_{\a\b\g}~.
\label{s-WeylW}
\eea
\end{subequations} 

It follows from \eqref{superPhi1} that 
\bea
\d_\s \Phi =0 ~\implies \d_\s \widehat{\Delta} \Phi = 3\s \widehat{\Delta} \Phi~.
\eea
In accordance with \eqref{superPsi1},  the super-Weyl transformation of $\Psi$ is 
\bea
\delta_\s \Psi = w \sigma  \Psi~. 
\eea
In follows from \eqref{lnPsi1} that the super-Weyl transformation of $\Delta \ln \bar\Psi$ is 
\begin{subequations}
\bea
\d_\s \Delta \ln \bar\Psi = 3\sigma \Delta \ln \bar\Psi~.
\eea 
By construction, we have  
\begin{align}
\delta_\sigma\,\widehat{\Delta} \ln\bar\Psi =& 3 \sigma  \widehat{\Delta}\ln\bar\Psi + w \widehat{\Delta} \bar\sigma~, 
\end{align}
and therefore
\begin{align}
\delta_\sigma \cQ =& 3 \sigma \cQ - \widehat{\Delta}\bar\sigma~.
\end{align}
\end{subequations}

Let us now turn to the $\cN=1$ super-Weyl anomalies  \cite{BPT,BK86}
In a classically super-Weyl invariant theory,  the effective action $\Gamma$ varies under the super-Weyl transformation as
\begin{align} 
\d_\sigma \G =& (a-c) \int \rd^4 x \,\rd^2 \q \,\cE \,\s W^{\a\b\g} W_{\a\b\g} 
+a \int 
 \rd^{4|4}z 
\,E \, \s( G^aG_a +2R \bar R)+{\rm c.c.}~,
\label{sWeyl-var}
\end{align}
modulo cohomologically trivial contributions. Here $ \rd^{4|4}z =  \rd^4 x \,\rd^2 \q \,\rd^2 \bar \q$,
$\cE$ is the chiral integration measure, 
and $E$ is the full superspace measure, see, e.g.,
\cite{KRTM1} for a review.
The coefficients $a$ and $c$ 
are of nontrivial significance in general (super)conformal 
field theories, see  \cite{Schwimmer:2010za, Komargodski:2011vj} and references therein.

It should be pointed  out that the functional (introduced at the linearised level in  \cite{FZ78}) 
\begin{align}
&\int \rd^4 x \,\rd^2 \q \,\cE \,W^{\a\b\g} W_{\a\b\g} 
+ \int 
 \rd^{4|4}z 
\,E \, (G^aG_a +2R \bar R) 
\end{align}
is a topological invariant, whose
real and imaginary parts correspond respectively to the usual Gauss-Bonnet
invariant and the Pontryagin term.
Its topological nature in superspace was first demonstrated in \cite{Buchbinder:1988tj}
(see also \cite{BK}).

Making use of  the identity 
\bea
\d_\s \int 
 \rd^{4|4}z 
\,E \,R \bar R = \frac 14 \int \rd^4 x \,\rd^2 \q \,\rd^2 \bar \q \,E \, \s \cD^2 R +{\rm c.c.}
\eea 
shows that  a local counterterm can be added to the effective action such that its super-Weyl variation \eqref{sWeyl-var} will turn into 
\begin{align} 
\d_\sigma \widetilde{\G }=& (a-c) \int \rd^4 x \,\rd^2 \q \,\cE \,\s W^{\a\b\g} W_{\a\b\g} 
+a \int 
 \rd^{4|4}z 
\,E \, \s \Big(G^aG_a +2R \bar R - \tfrac{1}{4} \cD^2 R\Big)+{\rm c.c.}\non \\
=& \int \rd^4 x \,\rd^2 \q \,\cE \, \s \Big\{ (a-c)W^{\a\b\g} W_{\a\b\g} + a \cQ\Big\}+{\rm c.c.}
\label{sWeyl-var2}
\end{align}

Now it is a short calculation to derive a Wess-Zumino action that leads to \eqref{sWeyl-var2}
\begin{align}
S_{\rm WZ} [\cD_A; \ln \Psi ] =& \frac{1}{w} \int \rd^4 x \,\rd^2 \q \,\cE \,\ln \Psi \Big\{ (a-c)W^{\a\b\g} W_{\a\b\g} + \frac{a}{w} \Delta \ln \bar \Psi \Big\} +{\rm c.c.}\non \\
&- \frac{a}{w^2} \int \rd^4 x \,\rd^2 \q \,\cE \,\ln \Psi \widehat{\Delta} \ln \bar \Psi~.
\end{align}
Its super-Weyl  variation coincides with  \eqref{sWeyl-var2},
\bea
\d_\s S_{\rm WZ} = \d_\sigma \widetilde{\G}~.
\eea

The Wess-Zumino action can be rewritten in a different form 
 \begin{align}
S_{\rm WZ} [\cD_A; \ln \Psi ] =& \frac{1}{w} \int \rd^4 x \,\rd^2 \q \,\cE \,\ln \Psi \Big\{ (a-c)W^{\a\b\g} W_{\a\b\g} + a \cQ \Big\} +{\rm c.c.}\non \\
&+ \frac{a}{16 w^2} \int  \rd^{4|4}z  \, E \,\Big\{ 
\cD^2 \ln \Psi \bar \cD^2 \ln \bar \Psi - 8\cD^\a \ln \Psi G_{\a\ad} \bar \cD^\ad \ln \bar \Psi \Big\}~.
\label{SWZN=1}
\end{align}
This action may be modified by adding a super-Weyl invariant functional of $\Psi $ and its conjugate.
The structure of \eqref{SWZN=1} is simpler than that of the Wess-Zumino action derived in \cite{Schwimmer:2010za}.\footnote{An alternative derivation 
of the Wess-Zumino action presented in \cite{Schwimmer:2010za} was given in \cite{Kuzenko:2019vvi}.}  
The latter contains cubic and quartic contributions 
in $\ln \Psi $ and $\ln \bar \Psi $.

It should be pointed out that integrating out the superfields $\Psi$ and $\bar \Psi$ in the action \eqref{SWZN=1}, 
one obtains the nonlocal action for the super-Weyl anomalies constructed in \cite{Butter:2013ura}. An alternative nonlocal effective action for the $\cN=1$ super-Weyl anomalies was computed in \cite{Buchbinder:1988yu}.


\section{Wess-Zumino action for $\cN = 2$ super Weyl anomalies} \label{Section4} 

To describe $\cN=2$ conformal supergravity, we will make use of the $\cN=1$ conformal superspace formulation developed by Butter \cite{ButterN=2}, see \cite{KRTM2} for a recent pedagogical review.

The $\cN=2$ superconformal analogue of the 
operator \eqref{Paneitz} was shown \cite{Butter:2013lta, deWit:2010za} to be the chiral projection operator
\begin{align} 
\bar \nabla^4:=\frac{1}{48} \bar \nabla^{ij} \bar \nabla_{ij}=-\frac{1}{48}
\bar \nabla^{\ad\bd} \bar \nabla_{\ad\bd}~, \quad \bar \nabla_{ij} := \bar{\nabla}_{\ad (i} \bar{\nabla}^\ad_{j)}~, \quad
\bar{\nabla}^{\ad \bd} := \bar{\nabla}^{(\ad}_i \bar{\nabla}^{\bd) i} ~.
\end{align}
The crucial property of $\bar \nabla^4$ is that its action on certain primary superfields results in primary ones.  
Specifically, let us consider a rank-$n$ spinor superfield $\f_{\a(n)} :=\f_{\a_1 \dots \a_n} =\f_{(\a_1 \dots \a_n)}$ that is $\sSU(2)_R$ neutral and has the following superconformal properties:
\bea
K^B  \phi_{\a(n)} = 0~, \quad {\mathbb D} \phi_{\a(n)} = (w-2)  \phi_{\a(n)}~, 
\quad {\mathbb Y}  \phi_{\a(n)} = 2(2-w)  \phi_{\a(n)}~.
\eea
Then its descendant 
\bea
 \Phi_{\a(n)} := \bar \nabla^4  \phi_{\a(n)}
 \eea
 is a primary covariantly chiral superfield,
 \begin{align}
K^B \Phi_{\a(n)} =0~, \quad \bar \nabla_j^\bd \Phi_{\a(n)} =0 \ ,\quad {\mathbb D} \Phi_{\a(n)} = w \Phi_{\a(n)} ~, \quad
{\mathbb Y} \Phi_{\a(n)} = -2w \Phi_{\a(n)} 
\label{chirals} 
\end{align}
Any primary covariantly chiral superfield is characterised by these properties. In particular, its $\sU_R$ charge is determined by its dimension.

Let ${\Psi}$ be a primary chiral scalar of non-zero dimension $w\in {\mathbb R} \setminus \{0\}$
 with the superconformal properties:
\begin{align}
	K^{B} {\Psi} = 0 ~, \qquad \bar \nabla^\ad_i {\Psi} = 0~, \qquad \mathbb{D} {\Psi} = w  \Psi 	~.
\end{align}
We assume $ \Psi $ to be nowhere vanishing, in particular $\Psi^{-1} $ exists. 
Then, it was shown in \cite{Butter:2013lta} that 
$\bar{\nabla}^4 \, \text{ln} \, \bar{\Psi}$ is a primary chiral superfield of dimension 2,
\begin{align}
	K^B \bar{\nabla}^4  \text{ln} \, \bar{\Psi} = 0 ~, \qquad \bar{\nabla}^{\ad}_i \bar{\nabla}^4  \text{ln} \, \bar{\Psi} = 0~, 
	\qquad \mathbb{D} \bar{\nabla}^4  \text{ln} \, \bar{\Psi} = 2 \bar{\nabla}^4  \text{ln} \, \bar{\Psi} ~.
\end{align}

Carrying out the degauging procedure from the superconformally covariant derivatives $\nabla_A$ to those for the so-called $\sSU(2)$ superspace
$\cD_A$ (see \cite{KRTM2} for the technical details), one obtains \cite{Butter:2013lta}
\begin{align}
		\bar{\nabla}^{4}  \ln \, \bar{\Psi} = \bar{\D} \, \text{ln} \, \bar{\Psi} + \frac{w}{2} \Big (  \bar{Y}^{\ad \bd} \bar{Y}_{\ad \bd} + \bar{S}_{ij} \bar{S}^{ij} 
	+ \frac 1 6 \bar{\mathcal{D}}_{ij} \bar{S}^{ij} \Big ) \equiv  \bar{\D} \, \text{ln} \, \bar{\Psi} + w 	\cQ
	 ~,
\end{align}
where $\bar{\D}$ denotes the chiral projecting operator of $\sSU(2)$ superspace
    \cite{KT-M09,Muller}
\bea
\bar{\D}
&=&\frac{1}{96} \Big((\cDB^{ij}+16\bar{S}^{ij})\cDB_{ij}
-(\cDB^{\ad\bd}-16\bar{Y}^{\ad\bd})\cDB_{\ad\bd} \Big)
\non\\
&=&\frac{1}{96} \Big(\cDB_{ij}(\cDB^{ij}+16\bar{S}^{ij})
-\cDB_{\ad\bd}(\cDB^{\ad\bd}-16\bar{Y}^{\ad\bd}) \Big)~,
\label{chiral-pr}
\eea
and $\bar{S}^{ij}$ and $\bar{Y}^{ \ad\bd}$ are torsion superfields of $\sSU(2)$ superspace.\footnote{A brief review of $\sSU(2)$ superspace is given in Appendix \ref{AppendixC}.}
Since
$\bar{\nabla}^{4} \, \text{ln} \, \bar{\Phi} $ and $\bar{\D} \, \text{ln} \, \bar{\Phi}$
 are both manifestly chiral, $\cQ$ is also chiral
\begin{align}
\bar{\mathcal{D}}^{\ad}_i \cQ = 0 ~.
\end{align}

Under the super-Weyl transformation, the following transformation laws hold:
\begin{subequations}
\begin{align} 
\d_\s \bar{\nabla}^{4} \, \text{ln} \, \bar{\Phi} =& 2\s \bar{\nabla}^{4} \, \text{ln} \, \bar{\Phi} ~,
\\
\d_\s  \bar{\D} \, \text{ln} \, \bar{\Phi} =& 2\s  \bar{\D} \, \text{ln} \, \bar{\Phi} +  \bar{\D} \s~.
\end{align}
Therefore, the super-Weyl transformation of $\cQ$ is 
\begin{align} 
\d_\s \cQ= 2 \s \cQ - w \s \cQ~.
\end{align}
\end{subequations}
 The interpretation of $\cQ$ 
as the $\cN=2$ supersymmetric $\cQ$-curvature was given in \cite{Levy:2019tjl}.

$\cN=2$ super-Weyl anomalies are described by the following super-Weyl  variation of the effective action \cite{K13}:
\bea
\d_\s \G = (a-c) \int \rd^4x\, \rd^4\q\, \cE\, \s W^{\a\b}W_{\a\b} 
-2a \int \rd^4x\, \rd^4\q\, \cE\, \s \cQ ~+~{\rm c.c.} ~,
\label{N=2sWeyl-var}
\eea
for some  anomaly coefficients $a$ and $c$. 
Here $\cE$ denotes the chiral density
(see, e.g.,  \cite{KT-M09} for more details), $W_{\a\b} =W_{\b\a} $  is the covariantly chiral
super-Weyl tensor.

Using the formalism described, it is not difficult to construct a Wess-Zumino action generating the super-Weyl variation \eqref{N=2sWeyl-var}. It is 
\bea
S_{\rm WZ}[\cD_A ; \Psi] &=& 
\frac{1}{w}\int \rd^4x\, \rd^4\q\, \cE\,  \Big\{  (a-c)\, W^{\a\b}W_{\a\b} 
-2 a\,  \cQ \Big\}\ln \Psi   ~+~{\rm c.c.} 
\non \\
&&  
-\frac{2a}{w^2} \int \rd^4 x \,{\rm d}^4\q\,{\rm d}^4{\bar \q}\,E\, \ln \Psi \ln \bar \Psi~.
\eea
Here $E$ is the full superspace integration density. Restricting $\Psi $ in $S_{\rm WZ}$ to be a reduced chiral superfield (which fixes $w=1$) leads to the action 
constructed in \cite{K13}.


\section{The logarithm construction in two dimensions}\label{Section5} 

In conclusion, we briefly elaborate on extending the construction advocated in this paper to two dimensions. The conformal $(p,q)$ superspace formalism 
in two dimensions was developed in \cite{Kuzenko:2022qnb}, with $p$ and $q$ non-negative integers.  
Conformal $(p,q)$ supergravity is described in curved superspace $\cM^{2|p,q)} $ with conformally flat geometry. Its covariant derivatives obey the anti-commutation relations 
\begin{align}
\{ \nabla_{+}^{\OI} , \nabla_{+}^{\OJ} \} =  2 \ri \d^{\OI \OJ} \nabla_{++} ~, \qquad  
\{ \nabla_{-}^{\UI} , \nabla_{-}^{\UJ} \} =  2 \ri \d^{\UI \UJ} \nabla_{--} ~,
\end{align}
with $\OI, \OJ=1,\dots , p$ and $\UI, \UJ=1,\dots, q$.\footnote{The case $p,q =0$ corresponds to conformal gravity. Here the conformally covariant derivatives commute, $[\nabla_{++}, \nabla_{--}] =0$.} 
In other words, $(p,q)$ conformal superspace has a flat superconformal connection. Non-trivial graded commutation relations for $\cN=(p,q)$ supergravity emerge upon degauging.

Wess-Zumino actions for $(p,q)$ super-Weyl anomalies may be derived by applying the $(p,q)$ logarithm constructions given in \cite{Kuzenko:2022qnb} for the cases $p,q\leq2$. In particular, given a nowhere vanishing primary scalar (super)field $\varphi$ of non-zero dimension $w$, one derives  the following primary descendants if $p,q \leq 1$:
\begin{subequations} 
	\begin{align}
		\cN = (0,0):& \qquad \nabla_{++} \nabla_{--} \, \text{ln} \, \varphi ~, \\
		\cN = (1,0):& \qquad \nabla_{--} \nabla_{+} \, \text{ln} \, \varphi ~, \\
		\cN = (1,1):& \qquad \hf [ \nabla_{+} , \nabla_{-} ] \, \text{ln} \, \varphi ~.
	\end{align}
\end{subequations}
These results are naturally extended to the cases $\cN=(2,0)$, $\cN=(2,1)$ and $\cN=(2,2)$  by working with a complex basis of spinor covariant derivatives, see 
 \cite{Kuzenko:2022qnb} for the technical details. 

\noindent
{\bf Acknowledgements:}\\ 
Discussions with Stefan Theisen are gratefully acknowledged. 
This work is supported in part by the Australian Research Council, project DP230101629.

\appendix 

\section{Elements of conformal gravity} \label{AppendixA}
This appendix is a shortened version of the reviews of conformal gravity included in \cite{BKNT-M1,BKNT, Kuzenko:2026vir}.

Conformal gravity in $d>2$ dimensions is obtained by gauging the conformal
algebra, $\mathfrak{so}(d,2)$, which is spanned by the translation ($P_a$), Lorentz ($M_{ab}$), dilatation ($\mathbb{D}$) and special conformal  ($K^a$) generators. Their non-vanishing commutation relations are
\begin{subequations}
	\label{ConAlg.1}
	\begin{align}
		[M_{ab} , M_{cd}] &= 2 \eta_{c[a} M_{b] d} - 2 \eta_{d [a} M_{b] c} ~, \\
		[M_{ab} , P_c ] &= 2 \eta_{c [a} P_{b]} ~, \quad [\mathbb D, P_a] = P_a ~, \\
		[M_{ab} , K_c] &= 2 \eta_{c[a} K_{b]}  ~, \quad  [\mathbb D, K_a] = - K_a ~, \\
		[K_a , P_b] &= 2 \eta_{ab} \mathbb D + 2 M_{ab} ~,
	\end{align}
\end{subequations}
with $\eta_{ab}$ the mostly plus Minkowski metric.
It is convenient to think of $\mathfrak{so}(d,2)$ as a sum of two Lie subalgebras 
generated by the operators
\begin{align}
	P_a \quad \& 
	\quad X_{\underline a} = (M_{ab},\mathbb{D},K^a)~,
	\label{generators}
\end{align}
respectively. Then, the conformal algebra \eqref{ConAlg.1} may be rewritten as follows
\bsubeq \label{structure}
\begin{align}
	[X_{\underline{a}} , X_{\underline{b}} ] &= -f_{\underline{a} \underline{b}}{}^{\underline{c}} X_{\underline{c}} \ , \\
	[X_{\underline{a}} , P_{{b}} ] &= -f_{\underline{a} { {b}}}{}^{\underline{c}} X_{\underline{c}}
	- f_{\underline{a} { {b}}}{}^{ {c}} P_{ {c}} \label{ConAlg.2}
	\ ,
\end{align}
\esubeq
where the structure constants can be read off from \eqref{ConAlg.1}.

Let  $\mathcal{M}^d$ be a manifolds admitting a spacetime metric.
To gauge the conformal
algebra $\mathfrak{so}(d,2)$ on $\cM^d$, we associate a connection one-form with each generator in \eqref{generators}, 
\begin{subequations} 
\bea
P_a & ~\to ~ &e^a = \rd x^m e_m{}^a~,  \\
X_{\underline{a}} = (M_{ab},\mathbb{D},K^a)&~\to ~&\o^{\underline{a}} = (\hat{\o}^{ab},\mathfrak{b},\mathfrak{f}_a)=\rd x^{m} \o_m{}^{\underline{a}}~, 
\eea 
\end{subequations}
and introduce a gauge covariant derivative $\nabla_a$  which replaces the translation generator $P_a$, 
\bea
P_a~\to ~\nabla_a =  e_a{}^m \partial_m  - \o_a{}^{\underline b} X_{\underline b} = e_a{}^m \partial_m - \hf \hat{\omega}_a{}^{bc} M_{bc} - {\mathfrak b}_a \mathbb D - \mathfrak{f}_{ab} K^b ~.
\eea
It is postulated that a commutator $[X_{\underline{a}} , \nabla_{{b}} ] $ is obtained from \eqref{ConAlg.2} by replacing $P_b \to \nabla_b$, 
\begin{align}
	[X_{\underline{a}} , \nabla_{{b}} ] &= -f_{\underline{a} { {b}}}{}^{\underline{c}} X_{\underline{c}}
	- f_{\underline{a} { {b}}}{}^{ {c}} \nabla_{ {c}} 
	\ .
\end{align}
The commutation relation $[P_a, P_b]=0$ turns into
\bea
[\nabla_a , \nabla_b] = -\cT_{ab}{}^c \nabla_c - \hf \cR(M)_{ab}{}^{cd} M_{cd}
- \cR(\mathbb{D})_{ab} \mathbb D - \cR(K)_{abc} K^c \ ,
\label{algebra}
\eea
where the right-hand side is given in terms of  the {torsion tensor} $\cT_{ab}{}^c $ and 
the {curvature tensors} $\cR(M)_{ab}{}^{cd}, ~
 \cR(\mathbb{D})_{ab}$ and $   \cR(K)_{abc}$. 

The gauge group of conformal gravity is generated by local transformations of the form
\begin{subequations}\label{CGgentransmations}
	\bea
	\delta_{\mathcal K} \nabla_a &=& [\mathcal{K},\nabla_a] \ , \qquad
	\mathcal{K} =
	\xi^b \nabla_b + \hf K^{bc} M_{bc} + \s \mathbb{D} + \L_b K^b ~.
	\eea
These gauge transformations act on a conformal tensor field $\mathcal{U}$ (with its indices suppressed) as 
\bea 
\label{CGgenMattertfs}
\d_{\mathcal K} \mathcal{U} = {\mathcal K} \mathcal{U} ~.
\eea
\end{subequations}
Of course, it is necessary to have a realisation of the conformal algebra on $\cU$. 
Such a realisation exists for conformal primary fields.

We say that $\mathcal{U}$ is a (conformal) primary field of dimension $w$ if the following conditions hold: 
(1)	$\cU$ is annihilated by the special
conformal generator, 
	\begin{subequations}
	\begin{align}
		K^a \mathcal{U} = 0~;
	\end{align}
(2)	$\cU$ is an eigenvector of the dilatation generator  with eigenvalue $w$,
	\begin{align}
		\mathbb D \cU = w \cU~.
	\end{align}
\end{subequations}
Thus 
\bea
\nabla_{a}\,\cU= \big(e_{a}{}^{m}\partial_m-\frac{1}{2} \hat{\omega}_{a}{}^{bc}M_{bc}-\D\mathfrak{b}_a\big) \cU~. 
\eea
Using this result, one may define the conformal transformation laws of descendants of $\cU$,  such as $\nabla_{a_1} \dots \nabla_{a_k} \cU$, for any positive integer $k$.

The geometric objects in \eqref{algebra} have the following explicit expressions:
\begin{subequations}
	\begin{align}
		\mathcal{T}_{ab}{}^c&=-\mathcal{C}_{ab}{}^{c}+2\hat{\o}_{[ab]}{}^c+2\mathfrak{b}_{[a}\delta_{b]}{}^{c}~,\\
		\mathcal{R}(M)_{ab}{}^{cd}&=\hat{R}_{ab}{}^{cd}+8\mathfrak{f}_{[a}{}^{[c}\delta_{b]}{}^{d]}~,\label{LorCur}\\ 
		\mathcal{R}(K)_{abc}&=-\mathcal{C}_{ab}{}^d\mathfrak{f}_{dc}-2\hat{\o}_{[a|c|}{}^{d}\mathfrak{f}_{b]d}-2{\mathfrak b}_{[a}\mathfrak{f}_{b]c}+2e_{[a}\mathfrak{f}_{b]c}~, \label{2.19cc}\\
		\mathcal{R}(\mathbb{D})_{ab}&= -\mathcal{C}_{ab}{}^{c}{\mathfrak b}_c+4\mathfrak{f}_{[ab]}+2e_{[a}{\mathfrak b}_{ b]}~.
	\end{align}
\end{subequations}
Here we have introduced the anholonomy coefficients $\mathcal{C}_{ab}{}^{c}$ defined by 
\begin{align}
	[e_a,e_b] = \mathcal{C}_{ab}{}^{c} e_c ~,
\end{align}
and the standard curvature tensor
\begin{align}
\hat{R}_{ab}{}^{cd}&=-\mathcal{C}_{ab}{}^{f}\hat{\o}_{f}{}^{cd}+2e_{[a}\hat{\o}_{b]}{}^{cd}
		-2\hat{\o}_{[a}{}^{cf}\hat{\o}_{b]f}{}^{d} \label{RiemannB}
\end{align}
constructed from the Lorentz connection $\hat{\omega}_a{}^{bc}$.  
Owing to its dependence on the dilatation connection, this curvature tensor does not satisfy the Bianchi identity $R_{[abc]d}=0$ unless $\mathfrak{b}_a=0$ 

To describe conformal gravity, conformally covariant constraints should be imposed on the torsion and curvature tensors \eqref{algebra}
such that the only independent geometric field, modulo gauge freedom, is the vielbein. Such constraints are:
\begin{subequations}
	\label{CC}
	\begin{align}
		\cT_{ab}{}^c &=0 ~, \label{CCa}\\
		\eta^{bd} \cR(M)_{abcd} &=0~. \label{CCb}
	\end{align}
\end{subequations}
The first constraint determines $\hat{\o}_{a}{}^{bc}$ in terms of the vielbein and dilatation connection  $\mathfrak{b}_a$, 
\begin{align}
  \hat{\omega}_{abc}=\o_{abc}-2\eta_{a[b}\mathfrak{b}_{c]}~,
  \quad 
  \o_{abc}
 =\frac{1}{2}\big(\mathcal{C}_{abc}-\mathcal{C}_{acb}-\mathcal{C}_{bca}\big)~,
 \label{LorentzConnection}
  \end{align}
where  $\o_{abc} \equiv \o_{abc} (e)$
 is the  torsion-free Lorentz connection. 
Constraint \eqref{CCa} is conformal since $K_a \cT_{bc}{}^d =0$, see \cite{BKNT-M1} for the technical details.
Constraint \eqref{CCb} 
determines the special conformal connection
$\mathfrak{f}_{a}{}^b$ in terms of $e^a$ and $\mathfrak{b}_a$,
 \begin{align}
 \mathfrak{f}_{ab}=-\frac{1}{2(d-2)}\hat{R}_{ab}+\frac{1}{4(d-1)(d-2)}\eta_{ab}\hat{R}~, 
 \end{align}
 where $\hat{R}_{ab}=\eta^{cd}\hat{R}_{acbd}$ is the (non-symmetric) Ricci tensor and $\hat{R}=\eta^{ab}\hat{R}_{ab}$ is the scalar curvature. 
Constraint \eqref{CCb} proves to be conformal, see   \cite{BKNT-M1} for the technical details.

Imposing the constraints \eqref{CC} leads to the algebra given by eq. \eqref{Cotton},
where $C_{abcd}$ is
the Weyl tensor satisfying the algebraic properties\footnote{
The symmetry property $C_{abcd} = C_{cdab}$ is not independent and follows from 
the others.} 
\bea
\eta^{bc} C_{abcd} =0\ , \quad
 C_{abcd} = C_{[ab][cd]} \ , \quad C_{[abc]d} = 0
\eea
and the Bianchi identity
\bea \nabla_{[a} C_{bc]}{}^{de} = - \frac{2}{d - 3} \nabla_{f} C_{[ab}{}^{f [d} \d_{c]}^{e]}  \ .
\eea
The Weyl tensor proves to be primary, $K^e  C_{abcd} =0$.\footnote{This implies that  it is independent of  the dilatation connection $\mathfrak{b}$. 
In other words, when the explicit expression for the Lorentz connection is used, 
all dependence on $\mathfrak{b}_a$ drops out of the Weyl tensor.}

Upon imposing the constraints \eqref{CC} we stay with two independent gauge fields, the vielbein $e^a$ and dilatation connection $\mathfrak{b}$. However, the latter  is a purely gauge degree of freedom since
\bea
\d_K(\L)\, \mathfrak{b}_a = -2 \L_a \ .
\non
\eea
Thus we can choose the gauge condition 
\bea
\mathfrak{b}_a=0~,
\label{b-gauge}
\eea
which fixes the special conformal gauge freedom. In this gauge,  the special conformal gauge freedom is fixed, and  the conformally covariant 
derivative takes the form
\bea
\nabla_a = \cD_a - \mathfrak{f}_a{}^b K_b \ , \quad \cD_a := e_a - \hf \omega_a{}^{bc} M_{bc} \ ,
\eea 
with $\cD_a$ the torsion-free Lorentz covariant derivative,
\begin{align}
	[\cD_a , \cD_b] 
	= - \hf R_{ab}{}^{cd} M_{cd} ~.
\end{align}
The special conformal connection takes the form
\bea 
\mathfrak{f}_{ab} = - \frac{1}{2 (d - 2)} R_{ab} + \frac{1}{4 (d - 1) (d - 2)} \eta_{ab} R  = - \hf P_{ab}~,
\eea
where
$R_{ac} := \eta^{bd} R_{abcd} $ is the Ricci tensor, $ R := \eta^{ab} R_{ab} $
the scalar curvature and $P_{ab}$  the {Schouten tensor}.
The Riemann tensor is
\bea  
 R_{ab}{}^{cd} = C_{ab}{}^{cd} +4 \d_{[a}^{[c} P_{b]}{}^{d]}  \ .
\eea

In the gauge \eqref{b-gauge},  two types of residual gauge transformations remain: 
\begin{itemize}
\item
Combined general coordinate and local Lorentz transformation 
\bea
	\delta_{\mathcal K} \cD_a = [\mathcal{K},\cD_a] \ , \qquad
	\mathcal{K} = \xi^b \cD_b + \hf K^{bc} M_{bc} 
\eea
acting on a tensor field $\cU$ (with indices suppressed) as
	\begin{align}
		\d_\cK \cU = \cK \cU ~.
	\end{align}	
\item 
Weyl transformation 
\begin{align}
	\mathcal{K}(\s) = \s \mathbb{D} + \frac{1}{2} \nabla_b \s K^b \quad \Longrightarrow \quad \d_{\mathcal{K}(\s)} \mathfrak{b}_a = 0
\end{align}
acting on a {primary field} $\cU$ of dimension $\D$ as 
{($\d_\s \equiv \d_{\mathcal{K}(\s) }$)}
\bea
\d_\s \cU = \D \s \cU~.
\eea
The Weyl transformation of the covariant derivative 
\begin{align}	
\d_\s \cD_a = \s \cD_a + \cD^b \s M_{ba} ~\implies ~
		\d_\s C_{abcd} = 2 \s C_{abcd} ~. 
		\label{WeylTr}
	\end{align}
\end{itemize}


\section{Elements of $\cN=1$ conformal superspace}\label{AppendixB} 

This appendix follows almost verbatim \cite{KRTM1, Kuzenko:2021qcx}.

Consider a $\cN=1$ superspace $\mathcal{M}^{4|4}$
parametrised by local coordinates 
$z^{M} = 
(x^{m},\theta^{\m},\bar \theta_{\dot{\mu}})$.  
The structure group is chosen to be $\sSU(2,2|1)$. The corresponding superalgebra is spanned by the translation $P_A=(P_a, Q_\a ,\bar Q^\ad)$, Lorentz $M_{ab}$, dilatation $\mathbb{D}$,  R-symmetry $\mathbb{Y}$ and the special conformal $K^A=(K^a, S^\a ,\bar S_\ad)$ generators. The covariant derivatives 
$\nabla_A$
then have the form
\begin{align}
	\nabla_A &= (\nabla_a, \nabla_\alpha, \bar\nabla^\ad)	=E_A{}^M \pa_M - \hf \Omega_A{}^{bc} M_{bc} - \ri \Phi_A \mathbb{Y}
	- B_A \mathbb{D} - \mathfrak{F}_{AB}K^B ~,
	\label{B.1}
\end{align}
where $E_{A}{}^{M}$ denotes the superspace inverse vielbein, $\Omega_A{}^{bc}$ the Lorentz connection,  $\Phi_A$  the  $\sU(1)_R$ connection, $B_A$
the dilatation connection, and  $\mathfrak F_{AB}$ the special
superconformal connection.

The graded commutation relations of $\nabla_A$ with the generators $X_{\underline a} = (M_{ab},\mathbb{D}, \mathbb{Y}, K^A)$ are obtained from those of $P_A$ with the same generators by replacing $P_A \to \nabla_A$. We recall that the general structure of the $\cN=1$ superconformal algebra is 
as follows:
\bsubeq\label{RigidAlgebra}
\begin{align}
[P_{ {A}} , P_{ {B}} \} &= -f_{{ {A}} { {B}}}{}^{{ {C}}} P_{ {C}}
	\ , \\
[X_{\underline{a}} , X_{\underline{b}} \} &= -f_{\underline{a} \underline{b}}{}^{\underline{c}} X_{\underline{c}} \ , \\
[X_{\underline{a}} , P_{{B}} \} &= -f_{\underline{a} { {B}}}{}^{\underline{c}} X_{\underline{c}}
	- f_{\underline{a} { {B}}}{}^{ {C}} P_{ {C}}
	\ . \label{mixing}
\end{align}
\esubeq
Here  $f_{{ {A}}{ {B}}}{}^{ {C}}$ contains only one non-zero component, which is $f_{\a}{}^\bd{}^c = 2 \ri \,(\s^c)_\a{}^\bd$.
 
For the reader's convenience, here we reproduce the relevant graded commutation relations.
The  $\sU(1)_R$, dilatation and special conformal generators obey
\begin{subequations}
	\begin{align}
		[\mathbb{Y}, \nabla_\a] &= \nabla_\a ~,\quad [\mathbb{Y}, \bar\nabla^\ad] = - \bar\nabla^\ad~,   \\
		[\mathbb{D}, \nabla_\aa] &= \nabla_\aa ~, \quad
		[\mathbb{D}, \nabla_\a] = \hf \nabla_\a ~, \quad
		[\mathbb{D}, \bar\nabla^\ad ] = \hf \bar\nabla^\ad ~\\
		[\mathbb{Y}, S^\a] &= - S^\a ~, \quad
		[\mathbb{Y}, \bar{S}_\ad] = \bar{S}_\ad~, \quad \{ S_\a , \bar{S}_\ad \} = 2 \ri  K_{\aa} \\
		[\mathbb{D}, K_\aa] &= - K_\aa ~, \quad
		[\mathbb{D}, S^\a] = - \hf S^\a~, \quad
		[\mathbb{D}, \bar{S}_\ad ] = - \hf \bar{S}_\ad ~,
	\end{align}
	while the algebra of $K_A$ and $\nabla_B$ takes the form
	\begin{align}
		[K_\aa, \nabla_\bb] &= 4 \big(\ve_{\ad \bd} M_{\a \b} +  \ve_{\a \b} \bar{M}_{\ad \bd} -  \ve_{\a \b} \ve_{\ad \bd} \mathbb{D} \big) ~, \\
		\{ S_\a , \nabla_\b \} &= \ve_{\a \b} \big( 2 \mathbb{D} - 3 \mathbb{Y} \big) - 4 M_{\a \b} ~, \\
		\{ \bar{S}_\ad , \bar{\nabla}_\bd \} &= - \ve_{\ad \bd} \big( 2 \mathbb{D} + 3 \mathbb{Y}) + 4 \bar{M}_{\ad \bd}  ~, \\
		[K_{\a \ad}, \nabla_\b] &= - 2 \ri \ve_{\a \b} \bar{S}_{\ad} \ , \qquad \qquad \qquad[K_\aa, \bar{\nabla}_\bd] =
		2 \ri  \ve_{\ad \bd} S_{\a} ~,  \\
		[S_\a , \nabla_\bb] &= 2 \ri \ve_{\a \b} \bar{\nabla}_{\bd} \ , \qquad \qquad \quad \qquad[\bar{S}_\ad , \nabla_\bb] =
		- 2 \ri \ve_{\ad \bd} \nabla_{\b} ~,
	\end{align}
\end{subequations}
where all other graded commutators vanish.

The covariant derivatives \eqref{B.1} obey the algebra
\begin{subequations}
	\label{CSSAlgebra}
	\bea
	\{ \nabla_{\a} , \nabla_{\b} \} & = & 0 ~, \quad \{\nabla_{\a} , \bar{\nabla}_{\ad} \} = - 2 \ri \nabla_{\a \ad} ~, \\
	\big[ \bar{\nabla}_{\ad} , \nabla_{\b \bd} \big] & = & - \ri \ve_{\ad \bd} \Big( 2 W_{\b}{}^{\g \d} M_{\g \d} + \frac{1}{2} \nabla^{\a} W_{\a \b \g} S^{\g} + \frac{1}{2} \nabla^{\a \gd} W_{\a \b}{}^{\g} K_{\g \gd} \Big) ~,
	\eea
\end{subequations}
Here $W_{\a \b \g}$ is the $\cN=1$ super-Weyl tensor and is subject to the constraints: 
\begin{subequations}
	\be
	K^D W_{\a \b \g} =0~, \quad \bar \nabla^\dd W_{\a\b\g}=0 ~, \quad 
	{\mathbb D} W_{\a\b\g} = \frac 32 W_{\a\b\g}~, \quad \mathbb{Y}W_{\a\b\g}=-W_{\a\b\g}~, 
	\ee
	as well as the Bianchi identity
	\be
	\label{B.3b}
	B_{\a\ad} :=  \ri \nabla^\b{}_{\ad} \nabla^\g W_{\a\b\g}
	=\ri \nabla_{\a}{}^{ \bd} \bar \nabla^\gd \bar W_{\ad\bd\gd}
	= \bar B_{\a\ad}~,
	\ee
\end{subequations}
where the primary superfield $B_{\a\ad}$ is the super-Bach tensor and was introduced in \cite{Buchbinder:1988tj}, see also \cite{KMT,KP,BK}.

We remind the reader that a tensor superfield $\J$ (with suppressed indices) is said to be primary 
and of dimension $w_\J$ and $\sU(1)_R$ charge $q_\J$ if the following conditions hold:
\bea
K^B \Psi = 0 ~, \qquad  \mathbb{D} {\Psi} = w_{{\Psi}} {\Psi} ~, \qquad 
\mathbb{Y} {\Psi} = q_{{\Psi}} {\Psi} ~.
\eea
Of particular importance are primary chiral superfields, which satisfy
\bea
K^B \Psi = 0 ~, \qquad \bar{\nabla}^{\bd} {\J} = 0~.
\eea
Requiring consistency of these constraints with the superconformal algebra yields highly non-trivial implications. Specifically, it must take the form $\Psi = \Psi_{\a(m)}$, and its $\sU(1)_R$ charge and dimension are related as follows:
\bea
\label{chiralDimChargeN=1}
q_\Psi = - \frac{2}{3} w_\Psi ~.
\eea

The chiral and full superspace integrals are related according to the rule
\bea
 \int \rd^4x\rd^2\theta\rd^2\bar\theta\, E\,  U
 =- \frac 14 \int \rd^4x \rd^2\theta\, \cE\, \bar \nabla^2 U~,\qquad E^{-1} = {\rm Ber}(E_A{}^M)~,
 \label{B.8}
\eea
where $U$ is a primary real  superfield of dimension $+2$.


\section{Elements of $\sSU(2)$ superspace} \label{AppendixC}

This appendix contains a brief review of the so-called $\sSU(2)$ superspace \cite{KLRT-M1}
which can be used to descrirbe $\cN=2$ conformal supergravity. 

The structure group of $\sSU(2)$ superspace is  $\sSL(2, \dsC) \times \sSU(2)$, 
and the covariant derivatives $\cD_A = (\cD_a, \cD_\a^i, \bar \cD^\ad_i)$ read
\bea
\cD_A &=& E_A - \Phi_A{}^{kl} J_{kl}- \hf \Omega_A{}^{bc} M_{bc} \non \\
		  &=& E_A - \Phi_A{}^{kl} J_{kl}- \Omega_A{}^{\b\g} M_{\b\g} 
		  - \bar{\Omega}_A{}^{ \dot{\b} \dot{\g} } \bar M_{\dot{\b}\dot{\g}}~.
\eea
Here $M_{cd}$ and $J_{kl}$ are the generators of the Lorentz and $\sSU(2)$ groups respectively, 
and $\O_A{}^{bc}$ and $\Phi_A{}^{kl}$ the corresponding connections. 
The action of the generators on the covariant derivatives are defined as:
\begin{subequations}
\bea
[M_{\a\b},  \cD_\g^i] &=& \ve_{\g (\a } \cD_{\b)}^i \ , \qquad [\bar M_{\ad \bd} , \bar \cD^i_\gd ] 
= \ve_{ \gd (\ad } \bar \cD^i_{\bd) i } ~, \\
\big[ J_{kl}, \cD_\a^i \big] &=& - \d^i_{(k} \cD_{\a l)} \ , \qquad 
[J_{kl}, \bar \cD^\ad_i] = - \ve_{i ( k} \bar \cD^\ad_{l)} \ .
\eea
\end{subequations}

The algebra of covariant derivatives is \cite{KLRT-M1}
\begin{subequations} \label{A.3}
\bea
\{\cD_\a^i,\cD_\b^j\}&=&\phantom{+}
4S^{ij}M_{\a\b}
+2\ve^{ij}\ve_{\a\b}Y^{\g\d}M_{\g\d}
+2\ve^{ij}\ve_{\a\b}\bar{W}^{\gd\dd}\bar{M}_{\gd\dd}
\non\\
&&
+2 \ve_{\a\b}\ve^{ij}S^{kl}J_{kl}
+4 Y_{\a\b}J^{ij}~,
\label{acr1} \\
\{\cDB^\ad_i,\cDB^\bd_j\}&=&
-4\bar{S}_{ij}\bar{M}^{\ad\bd}
-2\ve_{ij}\ve^{\ad\bd}\bar{Y}^{\gd\dd}\bar{M}_{\gd\dd}
-2\ve_{ij}\ve^{\ad\bd}{W}^{\g\d}M_{\g\d}
\non\\
&&
-2\ve_{ij}\ve^{\ad\bd}\bar{S}^{kl}J_{kl}
-4\bar{Y}^{\ad\bd}J_{ij}~,
\label{acr2} \\
\{\cD_\a^i,\cDB^\bd_j\}&=&
-2\ri\d^i_j(\s^c)_\a{}^\bd\cD_c
+4\d^{i}_{j}G^{\d\bd}M_{\a\d}
+4\d^{i}_{j}G_{\a\gd}\bar{M}^{\gd\bd}
+8 G_\a{}^\bd J^{i}{}_{j}~,
\label{acr3} 
\eea
\end{subequations}
see \cite{KLRT-M1} for the explicit expressions for  commutators
${[}\cD_a,\cD_\b^j{]}$ and ${[}\cD_a,\cDB^\bd_j{]}$.
Here the real four-vector $G_{\a \ad} $,
the complex symmetric  tensors $S^{ij}=S^{ji}$, $W_{\a\b}=W_{\b\a}$, 
$Y_{\a\b}=Y_{\b\a}$ and their complex conjugates 
$\bar{S}_{ij}:=\overline{S^{ij}}$, $\bar{W}_{\ad\bd}:=\overline{W_{\a\b}}$,
$\bar{Y}_{\ad\bd}:=\overline{Y_{\a\b}}$ 
are constrained by the Bianchi identities \cite{KLRT-M1} which we do not reproduce here.

The algebra  \eqref{A.3} is invariant 
under the super-Weyl transformations  \cite{KLRT-M1}
\begin{subequations}
\bea
\d_{\s} \cD_\a^i&=&\hf\sba\cD_\a^i+(\cD^{\g i}\s)M_{\g\a}-(\cD_{\a k}\s)J^{ki}~, \\
\d_{\s} \cDB_{\ad i}&=&\hf\s\cDB_{\ad i}+(\cDB^{\gd}_{i}\sba)\bar{M}_{\gd\ad}
+(\cDB_{\ad}^{k}\sba)J_{ki}~, 
\label{A.7b}
 \\
\d_{\s} \cD_a&=&
\hf(\s+\sba)\cD_a
+{\frac{\ri}4}(\s_a)^\a{}_{\bd}(\cD_{\a}^{ k}\s)\cDB^{\bd}_{ k}
+{\frac{\ri}4}(\s_a)^{\a}{}_\bd(\cDB^{\bd}_{ k}\sba)\cD_{\a}^{ k} \non \\
&&\quad -{\frac12}\big(\cD^b(\s+\sba)\big)M_{ab}
~, 
\eea
\end{subequations}
with  the parameter $\s$ being an arbitrary covariantly chiral superfield, ${\bar \cD}_{\ad i} \s=0$,
provided the  dimension-1 components of the torsion transform 
 as follows:
\begin{subequations}
\bea
\d_{\s} S^{ij}&=&\sba S^{ij}-{\frac14}\cD^{\g(i}\cD^{j)}_\g \s~, 
\label{super-Weyl-S} \\
\d_{\s} Y_{\a\b}&=&\sba Y_{\a\b}-{\frac14}\cD^{k}_{(\a}\cD_{\b)k}\s~,
\label{super-Weyl-Y} \\
\d_{\s} {W}_{\a \b}&=&\s {W}_{\a \b }~, \label{A.9c}\\
\d_{\s} G_{\a\bd} &=&
\hf(\s+\sba)G_{\a\bd} -{\frac{\ri}4}
\cD_{\a \bd} (\s-\sba)~.
 \label{super-Weyl-G} 
\eea
\end{subequations}
As is seen from \eqref{A.9c},
the covariantly chiral symmetric spinor $W_{\a\b}$ transforms homogeneously, 
and therefore it is a superfield extension of the Weyl tensor.

\begin{footnotesize}

\end{footnotesize}


\end{document}